\documentclass{aastex62}

\graphicspath{{./}{figures/}}
\received{3/4/2019}
\accepted{3/25/2019}
\submitjournal{ApJ Letters}

\shorttitle{Disk Mass Distributions across Protostellar Evolutionary Classes}
\shortauthors{Williams et al.}

\begin{document}

\title{The Ophiuchus DIsk Survey Employing ALMA (ODISEA): Disk Dust Mass Distributions across Protostellar Evolutionary Classes}

\correspondingauthor{Jonathan P. Williams}
\email{jw@hawaii.edu}

\author[0000-0001-5058-695X]{Jonathan P. Williams}
\affil{Institute for Astronomy,
University of Hawaii,
Honolulu, HI 96822, USA}

\author{Lucas Cieza}
\affiliation{Facultad de Ingenieria y Ciencias,
Nucleo de Astronomia,
Universidad Diego Portales,
Av. Ejercito 441. Santiago, Chile}
\nocollaboration

\author{Antonio Hales}
\affiliation{Joint ALMA Observatory,
Alonso de Cordova 3107,
Vitacura 763-0355, Santiago, Chile}
\nocollaboration

\author{Megan Ansdell}
\affiliation{Astronomy Department,
University of California,
Berkeley, CA 94720, USA}
\nocollaboration

\author{Dary Ruiz-Rodriguez}
\affiliation{Chester F. Carlson Center for Imaging Science,
Rochester Institute of Technology,
Rochester, NY 14623-5603, USA}
\nocollaboration

\author{Simon Casassus}
\affiliation{Departamento de Astronomia,
Universidad de Chile,
Casilla 36-D Santiago, Chile}
\nocollaboration

\author{Sebastian Perez}
\affiliation{Departamento de Astronomia,
Universidad de Chile,
Casilla 36-D Santiago, Chile}
\nocollaboration

\author{Alice Zurlo}
\affiliation{Facultad de Ingenieria y Ciencias,
Nucleo de Astronomia,
Universidad Diego Portales,
Av. Ejercito 441. Santiago, Chile}
\nocollaboration

\begin{abstract}
As protostars evolve from optically faint / infrared bright (Class I) sources
to optically bright / infrared faint (Class II) the solid material in their
surrounding disks accumulates into planetesimals and protoplanets.
The nearby, young Ophiuchus star-forming region contains hundreds of
protostars in a range of evolutionary states.
Using the Atacama Large Millimeter Array to observe their
millimeter continuum emission, we have measured masses of,
or placed strong upper limits on, the dust content of 279 disks.
The masses follow a log-normal distribution with a clear trend
of decreasing mass from less to more evolved protostellar infrared class.
The (logarithmic) mean Class I disk mass, $M=3.8\,M_\oplus$, is about 5 times greater
than the mean Class II disk mass, but the dispersion in each class is so high,
$\sigma_{{\rm log}\,M} \simeq 0.8-1$, that there is a large overlap
between the two distributions.
The disk mass distribution of flat-spectrum protostars lies in
between Classes I and II.
In addition, three Class III sources with little to no infrared
excess are detected with low disk masses, $M\simeq 0.3\,M_\oplus$.
Despite the clear trend of decreasing disk mass with protostellar
evolutionary state in this region, a comparison with surveys of Class II
disks in other regions shows that masses do not decrease monotonically
with age. This suggests that the cloud-scale environment may determine
the initial disk mass scale or that there is substantial dust regeneration
after 1\,Myr.

~
\end{abstract}


\section{Introduction} \label{sec:intro}
Stars form from the gravitational collapse of dense cores in
dusty molecular clouds.
As the obscuring dust is used up or otherwise dispersed,
the progression from embedded protostar to pre-main-sequence
star can be followed through the shift in the peak of the
spectral energy distribution (SED) from the far-infrared to optical.
The spectral slope from 2 to $25\,\mu$m provides the most commonly 
used classification, beginning with rising Class I, through Flat Spectrum,
to decreasing Class II, and finally Class III with weak or no infrared excesses.
The star has almost reached its final mass by the Class I phase,
within $\sim 10^5$ years after the onset of collapse
\citep{{2018A&A...618A.158K}},
but the remaining mass in the accompanying circumstellar disk and
how that evolves through the Classes as planets form over the next
several Myr is not known.

The material content in these small, cold, planet-forming disks is
best measured using high-resolution millimeter wavelength observations.
The Atacama Large Millimeter Array (ALMA)
has the imaging speed and sensitivity to carry out complete surveys
of disks in nearby star-forming regions, providing the essential
millimeter wavelength counterpart to previous infrared surveys.
However, the majority of ALMA disk measurements have been of Class II
sources since they are much more numerous than the short-lived Class I
phase and  very little material remains in the later Class III phase
\citep{2015A&A...583A..66H}.
To follow disk mass evolution requires a large survey of a
star-forming region with the right age to contain a mix of
many protostars in all evolutionary classes.

The Ophiuchus region is ideally suited on account of its
proximity, youth, and size
($\sim 140$\,pc, $\sim 1$\,Myr, and $\sim 300$ protostars).
Here, we discuss the Ophiuchus DIsk Survey Employing ALMA (ODISEA) project,
a complete survey of the millimeter continuum and CO line
emission from all the protostars identified in the \emph{Spitzer}
``Cores to Disks'' Legacy project \citep{2009ApJS..181..321E}.
The survey and first results are described in
\citet[hereafter Paper I]{2019MNRAS.482..698C}.
In this Letter, we describe the completion of the initial
survey and its vetting with the \emph{Gaia} mission.
We calculate disk masses and compare their distributions across
protostellar classes, revealing their evolution with more clarity
than before. We find that disk masses decline from Class I to Class II,
smoothly through the intermediate Flat Spectrum stage.
However, this seemingly simple picture of monotonic evolution is complexified
by a comparison with other regions, which shows that Ophiuchus Class II disks
have slightly, but significantly, lower masses than the slightly older
Lupus region. We conclude with a discussion of these results
and their implications for planet formation.

\section{Observations} \label{sec:obs}
The full set of 289 sources in the ODISEA sample was observed
in two samples, A and B, with 147 and 142 sources each.
Paper I describes the full sample selection and the observations of
sample A, consisting of Class I, Flat Spectrum, and bright
($K \leq 10$\,mag) Class II sources.
Here, we augment those data with the observations
of sample B, consisting of the fainter ($K > 10$\,mag)
Class II and Class III sources
in the same Cycle 4 ALMA program 2016.1.00545.S.

The observations of sample B were carried out in ALMA Band 6
with 40 antennas in the C40-3 array configuration
(15 to 500\,m baselines)
on 2018 May 2$^{\rm nd}$, and August 20$^{\rm th}$ and 21$^{\rm st}$.
The precipitable water vapor was 2.15, 0.95, and 1.43\,mm, respectively
with corresponding average system temperatures of 122, 98, and 109\,K.
The correlator was configured in the same way as for sample A,
with a total continuum bandwidth of 7.3 GHz centered at 225.4\,GHz
($\lambda = 1.33$\,mm).
There were also three higher spectral resolution windows
centered on the $J=2-1$ transition of CO, $^{13}$CO, and C$^{18}$O.
These lines constrain the disk gas content and will be discussed
in a future paper of the ODISEA series.

The visibilities were calibrated using the standard data pipeline
scripts using version 5.3.0 of the CASA software package.
The gain and bandpass calibrators were J1625-2527, J1517-2422, and J1924-2914.
The flux scale was referenced to J1733-1304 and J1517-2422
and has an uncertainty of 10\%.
Each of the 142 sources was observed for a total of 54 seconds.
Continuum images were created using the task {\tt tclean} and inspected
for multiplicity.
The beam size was similar for all sources, $\sim 0\farcs 98 \times 0\farcs 74$
at a position angle of $\sim 88^\circ$,
about four times larger than for sample A due to the more compact configuration.
The expectation (which was realized) was that the disk fluxes in this sample
would be lower due to their more evolved state and the lower stellar masses,
and the lower resolution was chosen to ensure high mass sensitivity
independent of disk size.
All but three sources were indeed unresolved and only one millimeter binary was found.
Photometry was carried out for the rest of the sample by fitting a point
source to the visibilities using the task {\tt uvmodelfit}.
Total fluxes for the binary and resolved sources were measured using
aperture photometry.

The distance to a source, $d$, used to be a considerable source of error in
determining disk and stellar properties but is now negligible thanks
to the \emph{Gaia} mission.
Out of 289 sources in the original ODISEA sample, 169 have parallaxes
($\pi$) in the Data Release 2 catalog \citep{2018A&A...616A...1G}.
23 sources (4 Class II, 19 Class III) have
$\pi < 2.5$\,mas which places them at a distance greater
than 400\,pc.
We considered these to be background objects, probably red giants,
and removed them from subsequent analysis
(all are undetected in our ALMA observations).
As in \citet{2018A&A...618L...3M},
we set $d = 1/\pi$ for those sources with $\pi/\sigma_\pi > 10$
since the inversion bias is very small in these cases
\citep{2015PASP..127..994B}.
For the 14 sources with larger fractional errors and the 106
with no measurement, we used a mean distance $\bar d = 139.4$\,pc.

Finally, we removed a misclassified Class III source, J162119.2-234229,
that was detected with a flux density $0.79\pm 0.13$\,mJy,
but is actually a Be star (HD\,147196).
The ALMA measurements and \emph{Gaia} distances for the final sample
of 265 sources
(279 disks after allowing for 12 binaries and one triple system)
is listed in Table 1.

\begin{deluxetable*}{rcrrrcc}[ht]
\tablecaption{Disk distances and flux densities\tablenotemark{a} \label{tab:fluxes}}
\tablenum{1}
\tablewidth{0pt}
\tablehead{
\colhead{Spitzer ID} &
\colhead{Class} & \colhead{$d$} &
\colhead{$F_{225GHz}$} & \colhead{$\sigma_{225GHz}$} &
\colhead{$\alpha_{2000}$} & \colhead{$\delta_{2000}$} \\
\colhead{(SSTc2d +)} & \colhead{} & \colhead{(pc)} &
\colhead{(mJy)} & \colhead{(mJy)} & 
\colhead{($^\circ$)} & \colhead{($^\circ$)} \\[-5mm]
}
\startdata
	J162034.2-242613 &  II & 139.40 &  0.22  & 0.11 &     ...    &      ...   \\
	J162118.5-225458 &  II & 138.97 &  3.26  & 0.13 &  245.32696 &  -22.91624 \\
	J162131.9-230140 &  II & 137.00 &  4.60  & 0.16 &  245.38302 &  -23.02800 \\
	J162138.7-225328 &   I & 139.40 &  0.17  & 0.20 &     ...    &      ...   \\
	J162142.0-231343 &  II & 138.94 &  1.75  & 0.11 &  245.42493 &  -23.22884 \\
\enddata
\tablenotetext{a}{Only the first 5 lines are shown here. The full table of
279 disks is available in machine-readable form in the online journal.}
\end{deluxetable*}

\section{Disk Mass Distributions} \label{sec:mass}
The dust masses of the disks are calculated in the simplest way,
under the assumption of optically thin emission throughout,
with a uniform temperature, $T_{\rm dust} = 20$\,K,
and opacity coefficient,
$\kappa_\nu = (\nu/100\,{\rm GHz})\,{\rm cm^2~g^{-1}}$.
For a disk with flux density $F_\nu$, this gives the standard formula,
\begin{equation}
M_{\rm dust} = \frac{F_\nu d^2}{\kappa_\nu B_\nu(T_{\rm dust})}
	     = 0.592\,M_\oplus
	       \left(\frac{F_{\rm 225 {\rm GHz}}}{1\,{\rm mJy}}\right)
	       \left(\frac{d}{140\,{\rm pc}}\right)^2,
\label{eq:kappa}
\end{equation}
where $B_\nu$ is the Planck function.
This approach is justified based on several reasons;
most sources in the survey only have one millimeter wavelength flux
measurement and are not well resolved;
detailed radiative transfer modeling of ALMA disk images around
protostars with similar luminosities as here find masses that are
very similar to those derived from this equation \citep{2017A&A...606A..88T};
it allows the cleanest comparison of disks both within the large Ophiuchus
sample and between surveys of other regions without adding uncertain
effects from poorly constrained parameter variations.
With that said, we return to the inherent assumptions
and possible complexities in the discussion of the results
in \S\ref{sec:discussion}.

A disk was deemed to be detected by ALMA if the measured flux density was
at least three times greater than the rms and the source position was within
$1''$ of the \emph{Spitzer} coordinates. 
For nondetections, we set a mass upper limit for a flux density threshold of
three times the rms.
Cumulative mass distributions were then determined using survival analysis,
which uses the constraints from upper limits in the data \citep{1985ApJ...293..192F}.
When plotted against the logarithm of the mass, we found that these
were well fit by the integral of a gaussian indicating that the underlying
probability distribution function is log-normal.
To aid the visualization of the distributions, we therefore show
both observed cumulative distributions with a $\pm 1\sigma$ range, and the
corresponding range of gaussian fits to the  probability distribution
in the figures below.

\subsection{Across Protostellar Evolutionary Classes} \label{subsec:YSOclasses}
\begin{figure}[ht]
	\plottwo{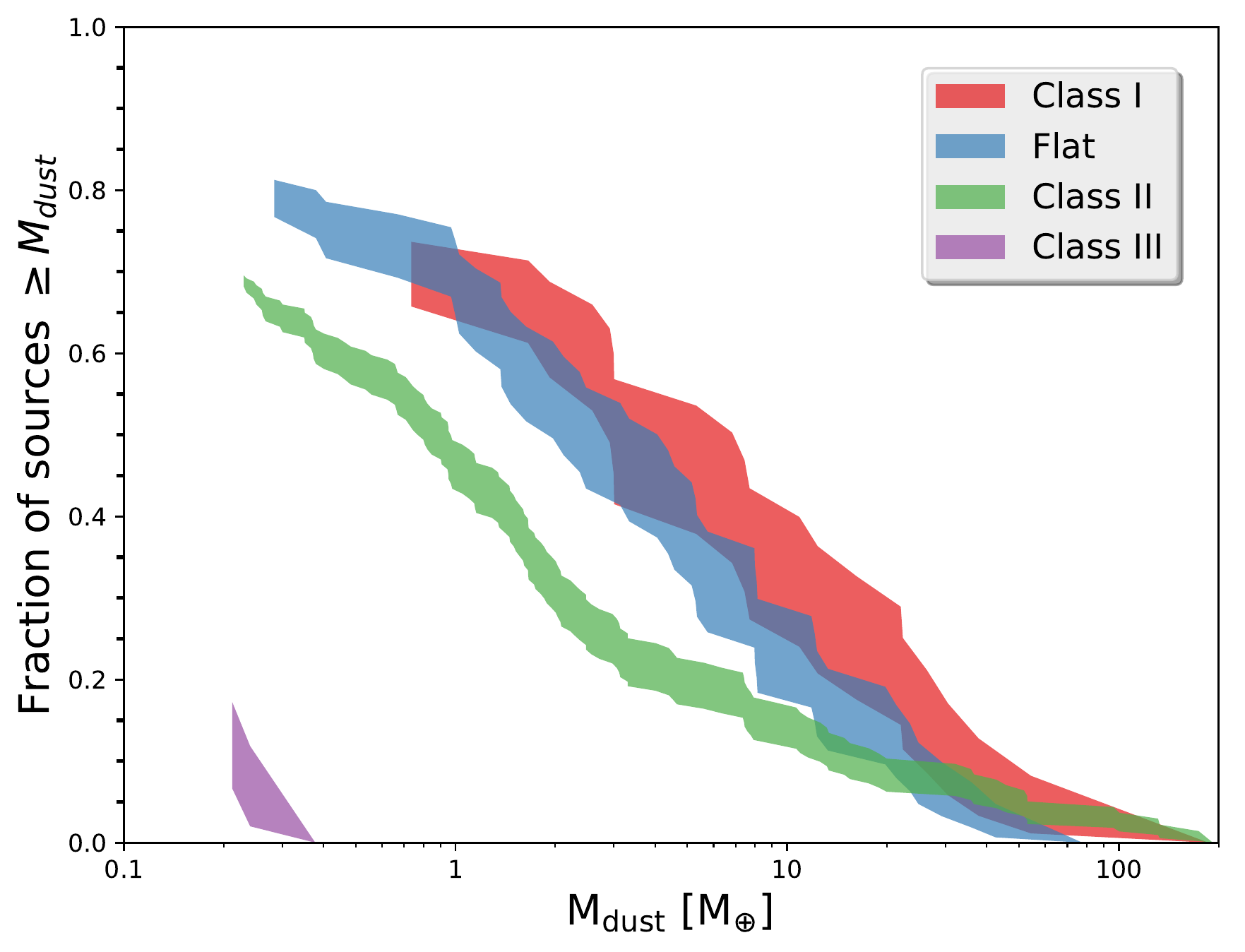}{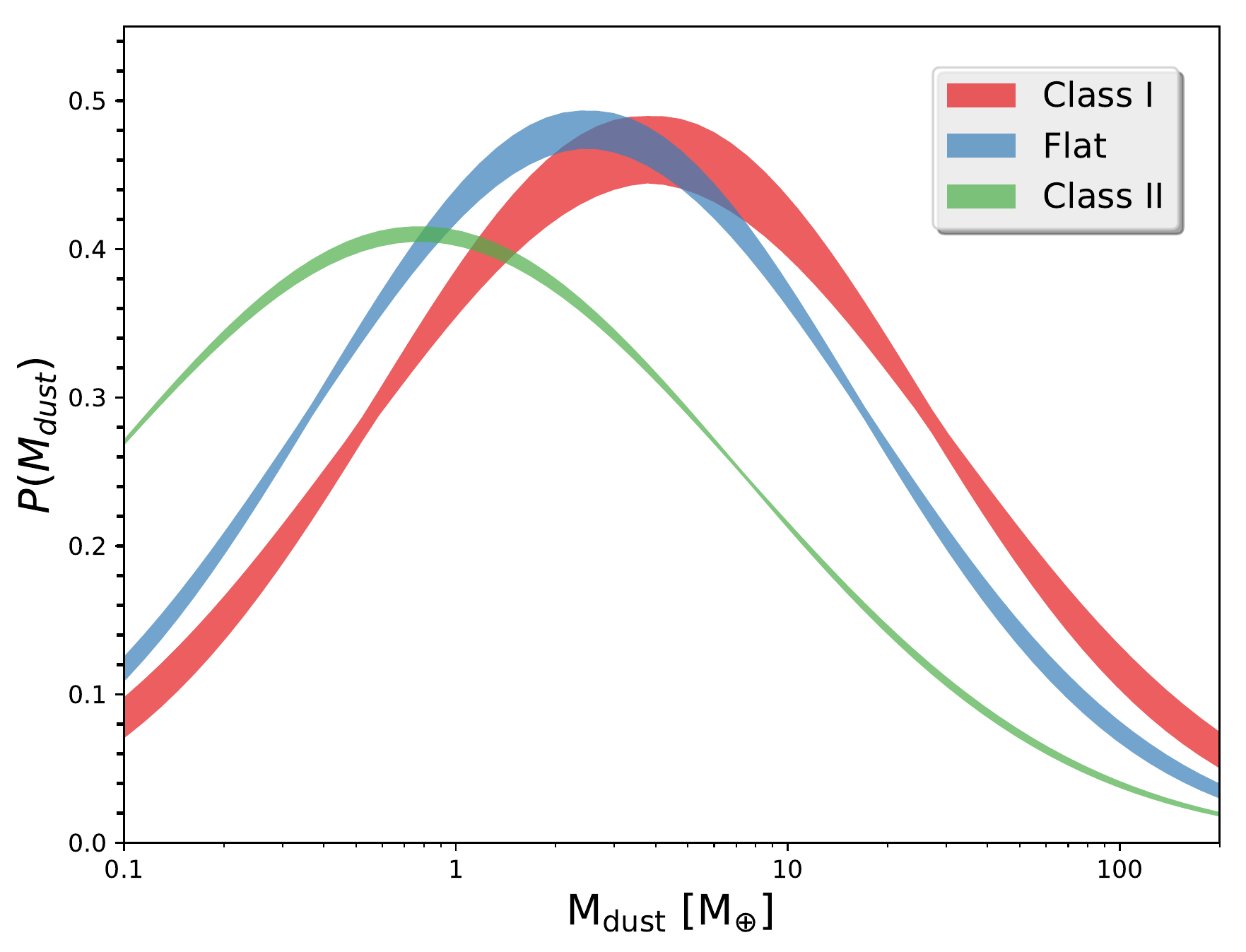}
	\caption{Dust mass distributions for Ophiuchus disks around protostars
	of different infrared evolutionary states. The cumulative distributions
	derived from the censored data are shown in the left hand plot,
	where the shading illustrates the $1\sigma$ uncertainty at each mass
	and the colors indicate protostellar class.
	The corresponding gaussian distributions for the probability distribution
	function for the Class I, Flat, and Class II sources is shown in the right
	panel, where the shading now illustrates the range of allowed fits.}
	\label{fig:YSOclasses}
\end{figure}

The dust mass distributions for each protostellar Class in Ophiuchus
are plotted in Figure~\ref{fig:YSOclasses}. For multiple systems, we have
only included the primary disk as defined by the infrared brightness
because the evolutionary state of the secondary (or tertiary) member is
not known from the {\emph Spitzer} data. 

The cumulative distributions shift to lower masses as protostars evolve.
However, the gaussian fits show that the mean Class I disk mass is only about
a factor of 5 higher than that of Class II, a difference that is less than the
standard deviation (Table~\ref{tab:YSOclasses}).
The upper quartile of Class II disks is more massive than the Class I mean
and the large overlap is clear in the plot of the fitted probability distributions.
Surprisingly, not all Class I or Flat Spectrum sources were
detected despite a $3\sigma$ mass sensitivity of $\sim 0.3\,M_\oplus$.
In contrast, we detected three Class III sources with dust masses $0.2-0.4\,M_\oplus$.
This is not enough to strongly constrain the full distribution but suggests
that more of these evolved disks may be detectable with a moderate increase
in sensitivity.

\begin{deluxetable}{crrr}[ht]
\tablecaption{Gaussian fit to Ophiuchus Disks \label{tab:YSOclasses}}
\tablecolumns{3}
\tablenum{2}
\tablewidth{0pt}
\tablehead{
\colhead{YSO Class} & \colhead{N} &
	\colhead{$\mu(M/M_\oplus)$\tablenotemark{a}}&
\colhead{$\sigma({\rm log_{10}} (M/M_\oplus))$}\\[-5mm]
}
\startdata
	I    &  28 & $3.83^{+1.62}_{-1.31}$ & $0.86^{+0.06}_{-0.02}$ \\[1mm]
	Flat &  50 & $2.49^{+0.82}_{-0.82}$ & $0.83^{+0.03}_{-0.01}$ \\[1mm]
	II   & 172 & $0.78^{+0.12}_{-0.11}$ & $0.97^{+0.06}_{-0.05}$ \\[1mm]
\enddata
	\tablenotetext{a}{The fits are for a gaussian in the logarithm of the mass with mean value $\log_{10}\mu$.}
\end{deluxetable}

Our results extend a recent ALMA survey of 49 Ophiuchus disks at $870\,\mu$m
that was weighted toward Class II sources but which found higher average flux
densities toward the small subset of Class I and Flat Spectrum objects in their
sample \citep{2017ApJ...851...83C}. They also found that binaries have significantly
lower flux densities. We have carried out high-resolution infrared imaging to
identify and study the effect of multiplicity on disks and will discuss this in a
future paper (Zurlo, A. et al. in preparation).
There are 40 binaries in our sample, a small fraction of the total,
and we found that omitting them from the disk distributions here did not
significantly change the log-normal fits.

\citet{2018ApJS..238...19T} found substantially higher disk masses for
Class 0 and I sources in Perseus based on VLA 9\,mm data.
However, there can be significant free-free emission at these long wavelengths
and the separation of the ionized gas and dust components is difficult.
A comparison of the two regions at millimeter and centimeter wavelengths
would help clarify the differences.

\citet{2019ApJ...873...54A} and \citet{2018ApJS..238...19T}
derived disk masses for Class I sources in Perseus and found values that
lie at the upper end of the Ophiuchus Class I distribution here
(when differences in the conversion from flux to mass are taken into account).
Their calculations require an extra step due, respectively, to lower
resolution 1\,mm data that blended disk and envelope emission and and
longer wavelength 9\,mm data that blended dust and free-free emission.
A complete, millimeter survey at sub-arcsecond resolution is required
to quantitatively compare the two regions.

\subsection{Across Star-Forming Regions} \label{subsec:other_regions}
ALMA has now surveyed the Class II disk population in several star-forming regions.
In Paper I, we found that the cumulative mass distribution for sample A
was similar to disks in the Taurus and Lupus regions.
However, about half of the Ophiuchus sources in that plot were Class I and
Flat Spectrum sources and we showed above that these tend to have more massive
disks than those around Class II.
Furthermore, the Class II sources in sample A are brighter in K-band
and the disk mass scales with stellar mass \citep{2013ApJ...771..129A}.
With the addition of sample B, we can now make a statistically fairer comparison
of complete populations of Class II sources.

\begin{figure}[ht]
	\plottwo{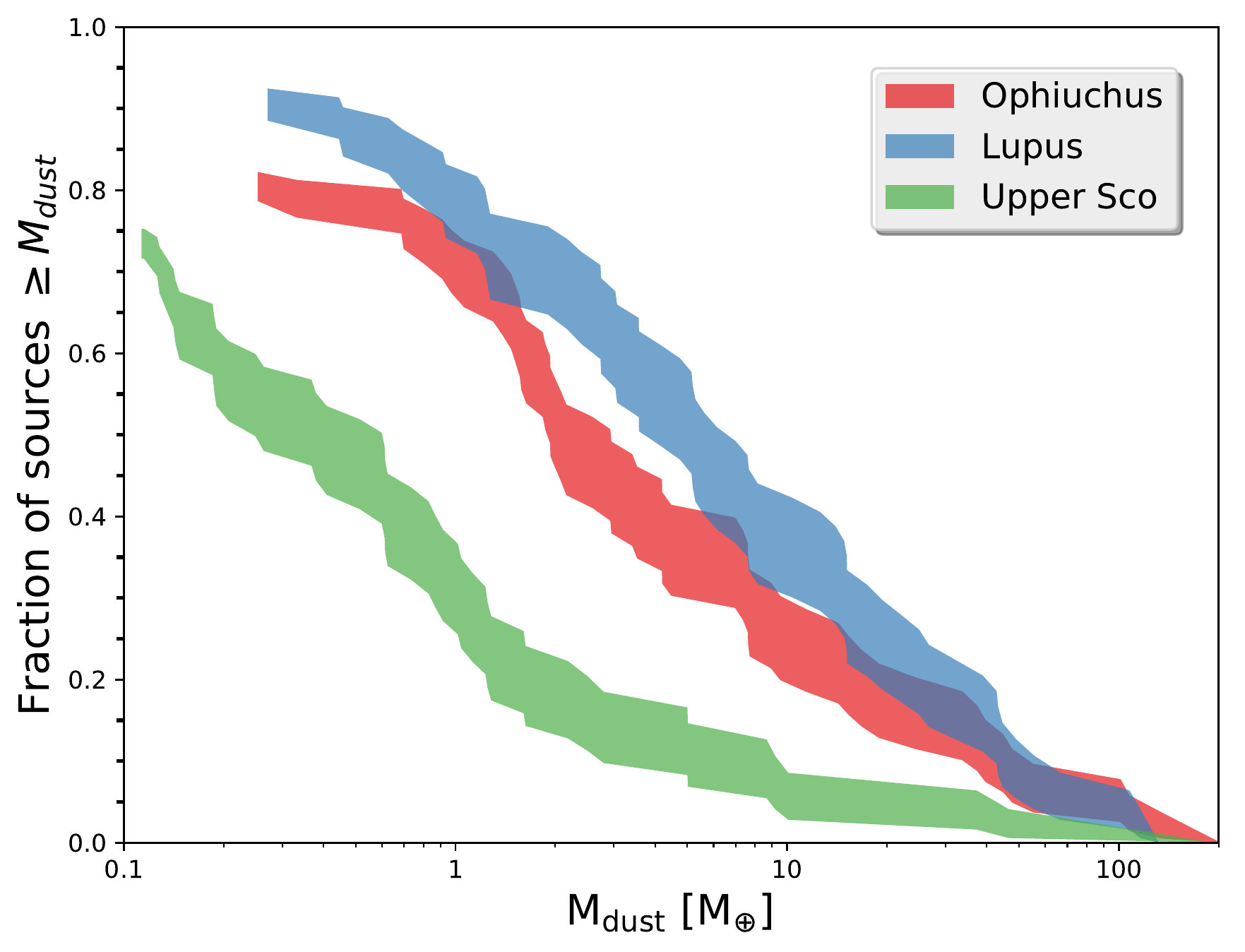}{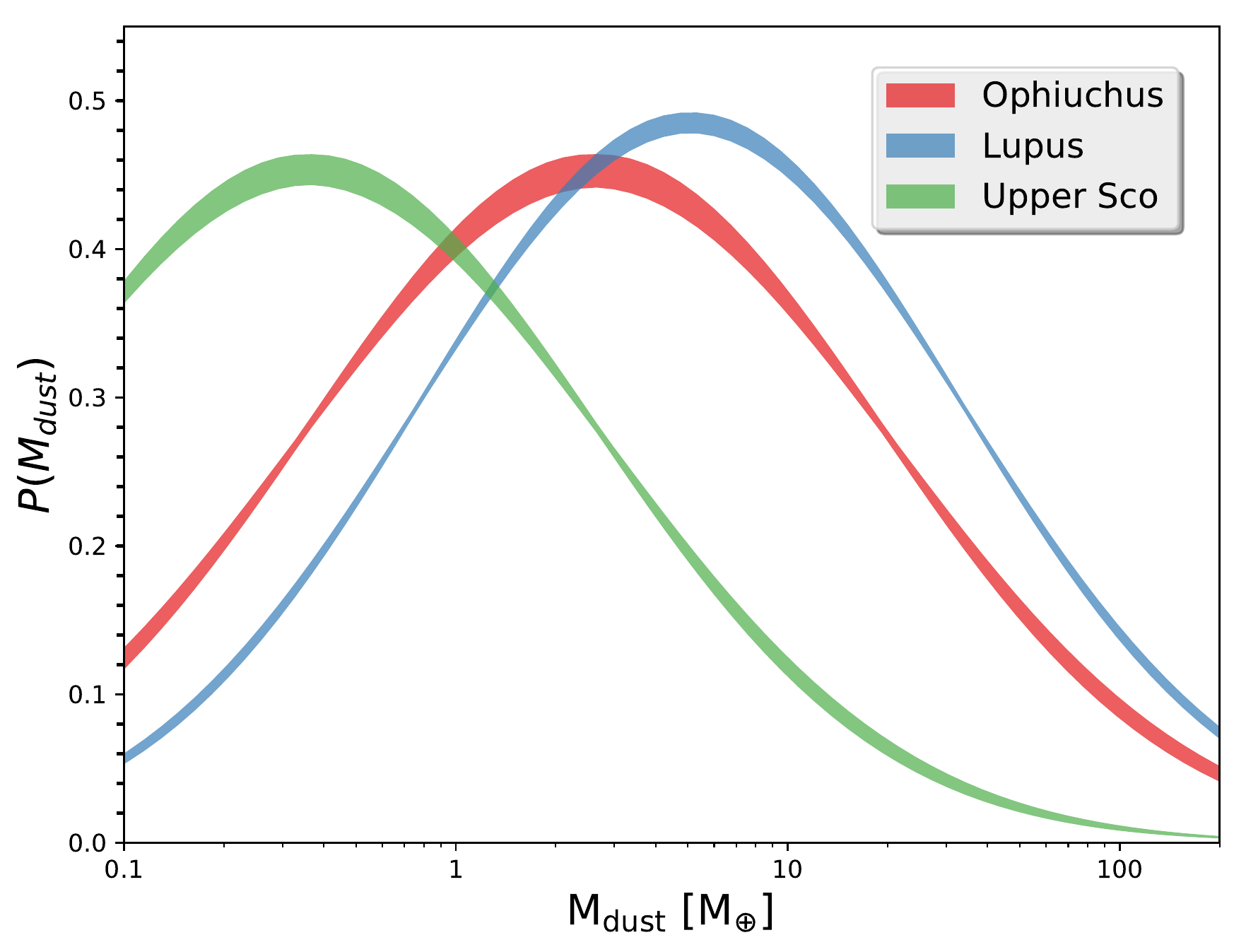}
	\caption{Dust mass distributions as in Figure~\ref{fig:YSOclasses},
	but now for disks around Class II protostars
	with estimated stellar masses $\gtrsim 0.1\,M_\odot$ in different
	star-forming regions.}
	\label{fig:regions}
\end{figure}

We show the Ophiuchus results in comparison with the ALMA surveys of
the Lupus and Upper Scorpius regions because these are similarly sensitive to
sub-Earth masses of dust and are complete or very nearly so.
Both regions are at a similar distance to Ophiuchus but are older,
$\sim 150$\,pc and $\sim 3$\,Myr for Lupus and $\sim 145$\,pc and $\sim 10$\,Myr
for Upper Scorpius \citep{2018ApJ...859...21A, 2016ApJ...827..142B}.
Disk masses were recalculated from the observed flux densities with the
same uniform temperature and opacity prescription as in Equation~\ref{eq:kappa}.
A proper comparison across regions requires accounting for possible differences
in the host stars.
Spectral types are not yet known for the full ODISEA sample
(the analysis of optical and infrared spectra will be presented in a future paper
in preparation by Ruiz-Rodriguez, D. et al. in preparation)
so we use a $1.2\,\mu$m J-band magnitude cutoff of 12 mag to restrict the
comparison to stars with masses estimated to be
$\gtrsim 0.2\,M_\odot$ in Figure~\ref{fig:regions}.

We now find that the Ophiuchus disks tend to have lower masses than those in Lupus.
The change is not surprising given the bias toward more massive disks in
sample A but the result upends the conventional wisdom
that disk masses decline monotonically with the age of the star-forming region.
The separation between the cumulative distributions is robust to different
J-band (or K-band) magnitude cutoffs and is discussed in more detail below.
Low disk masses were also recently reported in the similarly young Corona
Australis region \citep{2019arXiv190402409C}.
The results of the gaussian fits to the cumulative distributions from different
regions are provided in Table~\ref{tab:regions}.
The probability distributions are uniformly broad but the mean disk mass appears
to start low, then increase, before decreasing at late times.

\begin{deluxetable}{ccrr}[ht]
\tablecaption{Gaussian fits to Class II Disks in Different Regions\label{tab:regions}}
\tablecolumns{4}
\tablenum{3}
\tablewidth{0pt}
\tablehead{
\colhead{Region} &
\colhead{Age (Myr)} &
\colhead{$\mu(M/M_\oplus)$}&
\colhead{$\sigma({\rm log_{10}} (M/M_\oplus))$}\\
}
\startdata
	Ophiuchus        & $\sim 1$  & $2.62^{+0.83}_{-0.65}$ & $0.88^{+0.06}_{-0.04}$ \\[1mm]
	Lupus            & $\sim 3$  & $5.08^{+1.78}_{-1.41}$ & $0.82^{+0.01}_{-0.01}$ \\[1mm]
	Upper Scorpius   & $\sim 10$ & $0.36^{+0.10}_{-0.09}$ & $0.88^{+0.08}_{-0.05}$ \\[1mm]
\enddata
\end{deluxetable}

\section{Discussion} \label{sec:discussion}
The ODISEA project is the largest complete ALMA disk survey of a single
region to date and provides an unprecedented view of the disk mass
distribution across protostellar evolutionary states.
We find that disks are more massive in the early Class I stage,
and steadily decline through the Flat Spectrum to the Class II stage.
However, the difference in the mean mass from Class I to Class II
is only a factor of about 5 and the distributions have a large overlap.
In general, the dust masses are very low: only 10\% of Class I sources have
dust masses greater than the estimated total in the solar system, $30\,M_\oplus$
\citep{1981PThPS..70...35H};
and over half are less massive than $4\,M_\oplus$.
Based on exoplanet demographics, disks have a ``missing mass'' problem,
as noted on several occasions for Class II disks
\citep[most recently by][]{2018A&A...618L...3M},
and we now see that it extends to the young, embedded Class I phase.

We must therefore consider the validity of the assumptions about temperature,
opacity coefficient, and optical depth in \S\ref{sec:mass}.
The assumed temperature, 20\,K, is already low and does not leave much room
for increasing the mass. The beautiful high-resolution ALMA survey of
Class II disks by \citet{2018ApJ...869L..41A} shows that the optical depth
at 225\,GHz is generally less than one at $\sim 5$\,au scales
although there could be smaller, very high density concentrations
\citep{2018ApJ...865..157A}.
The opacity coefficient depends on size,
mineralogy, and shape of the dust grains but does not vary by more than
a factor of $\sim 3$ for all but the most extreme set of these parameters
\citep{1994ApJ...421..615P}.
However, any such flux-to-mass conversions are ultimately limited to constraints
on the mass of particles with sizes comparable to the observing wavelength and
there is ample reason to expect considerable mass in much larger bodies at early times:
for example, the existence of differentiated meteorites within $\sim 0.4$\,Myr
after the first solids in the solar system \citep{2017PNAS..114.6712K};
the detection of a planet in a Class II disk \citep{2016ApJ...826..206J};
and its implicit assumption in planetary population synthesis models
\citep{2018haex.bookE.143M}.

Despite their youth, the Ophiuchus Class II disks have slightly lower
masses than Class II disks in the older Lupus region.
A similar result was recently found for the comparably young
Corona Australis region \citep{2019arXiv190402409C}.
Based on the systematic decline in dust masses from $\sim 3$ to 10\,Myr
\citep{2017AJ....153..240A},
we would expect the Ophiuchus and Corona Australis disks to be more,
not less, massive.
Because disk masses also correlate with stellar mass, such comparisons between
regions consider possible differences in the stellar sample.
The stellar initial mass function is known to be quite
universal \citep{2014prpl.conf...53O} although there may be differences
around the substellar boundary. We have used a simple near-infrared luminosity
cutoff to stay above that limit and will explore this further as we learn more
about the stellar properties across the full sample.

The inherently limited sample sizes restrict the statistical rigor with which
evolutionary trends can be dissected by stellar mass and other parameters,
but the low disk masses in the $\sim 1$\,Myr old Ophiuchus and Corona Australis 
regions appear to be an inherent trait.
One possibility is simply that disk masses depend on the local (cloud)
environment where the stars form. 
Such differences do not, however, appear to
manifest themselves in stellar properties such as mass distribution or binarity.
A more speculative, though exciting, alternative is that the young Class II
disks in Ophiuchus and Corona Australis are indeed protoplanetary with most
of the mass in planetesimal and smaller sizes, and that the slightly older
Class II disks in Lupus and Taurus might be closer to the peak of planet formation
with Earth masses of second-generation dust produced as the disk is stirred up due to
the aggregation of planetesimals into protoplanets.

Demographic studies such as these provide useful insights into
disk evolution and planet formation.
Nevertheless, as with exoplanet observations, it is important to go beyond
a single measure of an object and to gather more information.
An important next step will be higher resolution observations to measure sizes
and structure, and to see how these vary with protostellar class, stellar mass,
and from region to region.

\acknowledgments
J.P.W. thanks Ewine van Dishoeck for comments.
L.C. was supported by CONICYT-FONDECYT grant number 1171246.
A.Z. acknowledges support from the CONICYT + PAI/ Convocatoria
nacional subvenci\'on a la instalaci\'on en la academia,
convocatoria 2017 + Folio PAI77170087.
This paper makes use of the following ALMA data: ADS/JAO.ALMA
\#2016.1.00545.S. ALMA is a partnership of ESO
(representing its member states), NSF (USA), and NINS (Japan),
together with NRC (Canada), NSC and ASIAA (Taiwan),
and KASI (Republic of Korea), in cooperation with the Republic of Chile.
The Joint ALMA  Observatory is operated by ESO, AUI/NRAO, and NAOJ.
The National Radio Astronomy Observatory is a facility of the National
Science Foundation operated under cooperative agreement by
Associated Universities, Inc.

\vspace{5mm}
\facilities{ALMA}
\software{astropy \citep{2013A&A...558A..33A}}



\end{document}